\documentclass[11pt]{article}

\newcommand{\ee}{ E^{\{1\}} }
\newcommand{\Ee}{ E^{\{2\}} }
\newcommand{\Ph}{ \phi^{\{2\}} }
\newcommand{\ps}{ \psi^{\{1\}} }
\newcommand{\Ps}{ \psi^{\{2\}} }
\newcommand{\ph}{ \phi^{\{1\}} }

\usepackage{geometry}  
\usepackage[parfill]{parskip}  
\usepackage{graphicx}
\usepackage{amsmath, amsthm, amsfonts}
\usepackage{lipsum}
\usepackage[utf8]{inputenc}
\usepackage[T1]{fontenc}

\newcommand{\sect}[1]{\setcounter{equation}{0}\section{#1}}

\usepackage{color}
     \usepackage[colorlinks]{hyperref}
\hypersetup{linkcolor=blue,%
citecolor=red,%
urlcolor=cyan}
\usepackage{url} 
\usepackage{epstopdf}
\usepackage{amssymb}

\title{SUSY partners and $S$-matrix poles of the one dimensional
\\ Rosen--Morse II  potential}

\author{ 
Carlos San Mill\'an$^a$
\footnote{carlos.san-millan@uva.es, ORCID: 
\href{https://orcid.org/0000-0001-7506-5552}{0000-0001-7506-5552}}\,,
Manuel Gadella$^a$
\footnote{manuelgadella1@gmail.com, ORCID: 
\href{https://orcid.org/0000-0001-8860-990X}{0000-0001-8860-990X}}\,,
\c{S}eng\"ul Kuru$^b$
\footnote{kuru@science.ankara.edu.tr, ORCID: 
\href{https://orcid.org/0000-0001-6380-280X}{0000-0001-6380-280X}}\,,
Javier Negro$^a$
\footnote{jnegro@fta.uva.es, ORCID: 
\href{http://orcid.org/0000-0002-0847-6420}{0000-0002-0847-6420}}
\bigskip
\\
\noindent
$^a$\,Departamento de F\'{\i}sica Te\'orica, At\'omica y
\'Optica, and IMUVA,\\ Universidad de Valladolid,  47011 Valladolid, Spain
\\ 
 \noindent
$^b$\,Department of Physics, Faculty of Science, Ankara University, 06100 Ankara, T\"urkiye}

\begin{document}
\maketitle

\begin{abstract}
Among the list of one dimensional solvable Hamiltonians, we find the Hamiltonian with the Rosen--Morse II potential. The first objective is to analyze the scattering matrix corresponding to this potential. We show that it includes a series of poles corresponding to the types of redundant poles or anti-bound poles. In some cases, there are even bound states and this depends on the values of given parameters. Then, we perform different supersymmetric transformations on the original Hamiltonian either using the ground state (for those situations where there are bound states) wave functions, or other solutions that come from anti-bound states or redundant states. We study the properties of these transformations. 

\end{abstract}

\sect{Introduction} \label{I1}

This is a new contribution to the study of solvable one dimensional potentials from two different points of view. First of all from the properties of the scattering matrix and the analysis of its poles. Second from the perspective of the successive supersymmetric (SUSY) transformations \cite{CKS} of the original Hamiltonian with respect to different solutions of the Schr\"odinger equation with energies given by the different poles of the scattering matrix. We shall show that such solutions associated to the $S$-matrix poles are particularly suitable to build SUSY transformations with many interesting properties.

Here, the object of our study is focused on the Rosen--Morse II potential, originally  developed to describe  vibration of molecules \cite{rosen32}. This potential, and its Hamiltonian, has been studied in previous articles from different points of view \cite{MMN,LM,QUESNE,RGKN,hussin19,hussin21}. We intend to continue using the Rosen--Morse II potential, in the same spirit we have used with other solvable one dimensional Hamiltonians such as the P\"oschl-Teller potential \cite{RGKN,CG,CGKN}, the hyperbolic step potential \cite{GKN}, or the Morse potential \cite{GAKN}. In the latter, we have found an interesting structure in the scattering matrix. It includes a series of bound and anti-bound simple poles and two series of redundant poles \cite{MA,BISWAS}. On each series, the   solutions of the Schr\"odinger matrix with energies located at the poles of $S(E)$ are related via first order differential ladder operators. These type of ladder operators relate the Gamow wave functions for the resonance poles on the scattering matrix for the hyperbolic P\"oschl-Teller potential \cite{CGKN}. 

In the present paper, we make it explicit the scattering matrix structure of the one dimensional Hamiltonian with Rosen--Morse II potential, which can be exactly obtained. We show the existence of redundant poles and anti-bound poles and, depending on the shape of this potential which, at its turn, depends on the values of certain parameters, the possible existence of bound states. Unfortunately, first order ladder differential operators cannot exist in this case; but there are first order ``shift operators'' responsible for the shape invariance, which are proven to be quite useful. Also, we observe the absence of resonance poles. This discussion is focused in Section 2.

In the second half of this paper, Sections 3 and 4, we consider the SUSY transformed Hamiltonians after the original Rosen--Morse II. Here, we use different seed functions. The simplest situation is derived from the use of the ground state of the Rosen--Morse II Hamiltonian as a seed function. Then, a whole sequence of SUSY partner Hamiltonians may be explicitly derived. If we depart from an original Hamiltonian which has $N$ bound states, for each of the successive SUSY transformation performed, we loose one bound state. This process stops at the $N$-th transformation. The resulting Hamiltonian and all Hamiltonians resulting after successive transformations does not have bound states. Nevertheless, the number of redundant poles, which is finite, and the number of  anti-bound poles, which is infinite, remain the same.

There are two other possibilities of using other types of seed functions. Should we use one wave function for a redundant state as seed function, the number of poles of $S(E)$ of each kind would be fixed. If this seed function were the wave function for an anti-bound pole (anti-bound state), then a new bound state appears for each transformation. Then, we can say that the anti-bound state becomes a bound state under a SUSY transformation.  In any case,  the number of redundant poles remains the same. 

We finish the present manuscript with a discussion on the equivalence of some SUSY transformations on the Rosen--Morse II Hamiltonians, in Section 4 and some concluding remarks on the applications of our study. 

\sect{The Rosen--Morse II potential}

To begin with,  let us write the one dimensional Hamiltonian 
\[
H_\lambda:= -\frac{d^2}{dx^2} + V_\lambda(x), 
\]
where the potential $V_\lambda(x)$ is given by (we have chosen $\hbar^2/(2m)=1$)
\begin{equation}\label{1}
V_\lambda(x) : = -\left(\lambda^2-\frac{1}{4}\right)\operatorname{sech}^2(x)+2\beta \operatorname{tanh(x)}\,.
\end{equation}
with $\lambda$ and $\beta$  parameters whose values will be fixed later. Potential \eqref{1} is known as the {\it Rosen--Morse II}
 potential \cite{CKS}. In the limit $\beta =0$, it  becomes the hyperbolic 
 P\"oschl--Teller potential. Thus, the parameter $\beta$ is responsible for the asymmetry of the potential. Note that the transformation $\beta \longmapsto -\beta$ is equivalent to the transformation given by the parity operator $x \longmapsto -x$, so that we may take $\beta >0$ without loss of generality. 
In addition,  there is the reflection symmetry $\lambda \to -\lambda$:  $H_\lambda=H_{-\lambda}$, which will be present in   all the subsequent discussion. Henceforth, we will assume that $\lambda>0$, with no loss of generality. 
This potential has different shapes depending on certain relations of the two parameters $\beta$ and $\lambda$, as may be seen in Figure 1 for some particular values. The situations to be considered are three:

\begin{itemize}

\item{The potential has a minimum and, therefore has a potential well shape if $0<\beta<\left(\lambda^2-\frac{1}{4}\right)$. This is shown by the black curve in Figure 1.}

\item{A typical step potential is shown provided that $0<\left(\lambda^2-\frac{1}{4}\right)<\beta$. See blue dashing curve in Figure 1.}

    \item{
   A bump on  the step (or asymmetric barrier) appears for values of the form $\lambda = i \ell$, $\ell\in \mathbb{R}^+$, with $ \left|-\ell^2-\frac{1}{4}\right|>\beta 
    > 0$  as shown in red dotted curve in Figure 1. }

\end{itemize}

\begin{figure}
		\centering
		\includegraphics[width=0.5\textwidth]{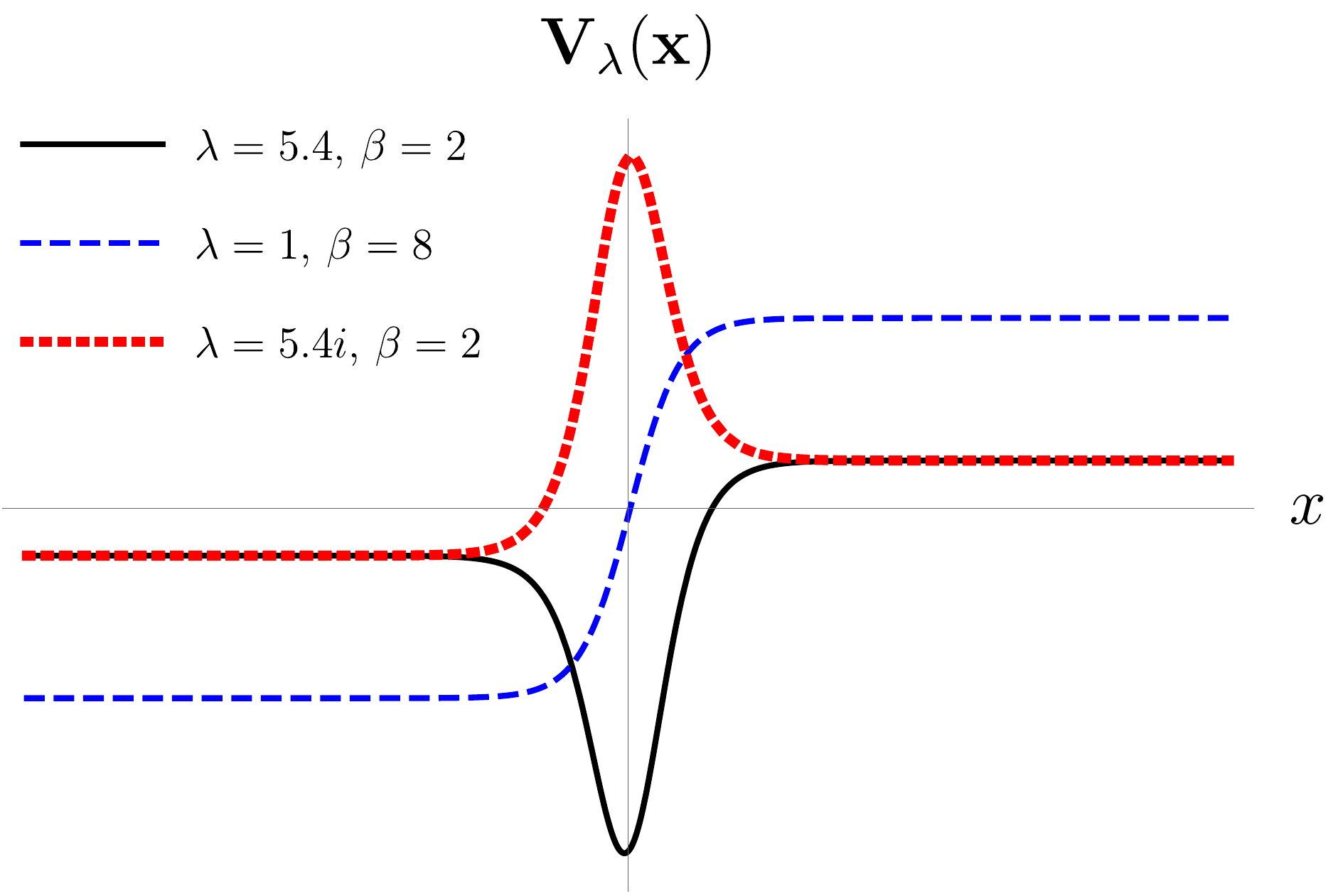}
		\caption{Rosen--Morse II potential for  distinct cases considered: black (continuous) for potential well, blue (dashing) for step potential, and red (dotted) 
		asymmetric barrier. }
			\label{fig:V0}
	\end{figure}

\subsection{On the scattering matrix}

Next, we intend to obtain scattering features given by the $S$-matrix poles corresponding to the Hamiltonian pair $\{H_0,H_\lambda\}$, with $H_0=-d^2/dx^2$ and $H_\lambda = H_0 +V_\lambda(x)$. First of all, we solve the Schr\"odinger equation $H_\lambda \, \Psi(x) = E \Psi(x)$, which is exactly solvable, provided that we perform a series of changes in both the solution function and the independent variable. We begin with the following changes, which include the introduction of two new parameters $k$ and $k'$, which we give the respective meaning of  income and outcome moments:

\begin{equation}\label{2}
\displaystyle \Psi(z)=(1-z)^{ik'/2}(1+z)^{ik/2}f(z),\quad	z=\tanh x\,, \quad
 k=\sqrt{E+2\beta},\quad  k'=\sqrt{E-2\beta}\,.
 \end{equation}

This gives the following equation (appropriate to describe Jacobi polynomials for some specific values of the parameters)
\begin{eqnarray}\label{3}
(1-z^2)f''(z)+\big[ik-ik'-(2+ik+ik')z\big]f'(z) + 
\\[2ex]\nonumber
\left[ \left( \lambda -\frac 12 - \frac{ik}{2} - \frac{ik'}{2} \right)   
\times  
\left( \frac{i k}{2} +\frac{i k'}{2}+\frac{1}{2}+\lambda \right)  
\right] \, f(z)=0\,.
\end{eqnarray}
We are looking for exact solutions of \eqref{3} for any value of $E$, $\lambda$ and $\beta$. Observe that from (\ref{2}),
\begin{equation}\label{5}
E = \frac{k^2+k'^2}{2}\,, \qquad \beta = \frac{k^2-k'^2}{4}\,.
\end{equation}
This identity has particular importance as we shall see later.  Then, on the resulting equation, we perform a new change in the variable given by

\begin{equation}\label{6}
z=2t-1\,,
\end{equation}
so that the Schr\"odinger equation is transformed into the following second order differential equation  \cite{OLVER}:
\begin{eqnarray}\label{7}
(1-t)t f''(t)+\big[1+ik-(2+ik+ik')t\big]f'(t) 
-\left(\frac{i k}{2} +\frac{i k'}{2}+\frac{1}{2}-\lambda\right)\left(\frac{i k}{2} +\frac{i k'}{2}+\frac{1}{2}+\lambda\right)f(t)=0 \,.
\end{eqnarray}	

This is a hypergeometric equation, for which the solutions are well known \cite{OLVER}. Once we have the general solution of \eqref{7}, we obtain the general solution of the original Schr\"odinger equation by reverting the transformations given in \eqref{6} and \eqref{2}. This is:
\begin{equation}\label{8}
\begin{array}{l}
\Psi(x)=A\psi(x)+B\phi(x) 
\\[2ex]  
=A\, (1-\tanh (x))^{ik'/2} (1+\tanh (x))^{ik/2} 
 {}_2F_1\left(\begin{matrix}
			\frac{1}{2}-\lambda+i\frac{k+k'}{2}\,, \frac{1}{2}+\lambda+i\frac{k+k'}{2}
			\\[1ex]  
			1+ik\end{matrix}\, ; \frac{1+\tanh(x)}{2}\right) 
			\\[3ex] 
			+ B\,2^{ik}\, (1-\tanh (x))^{ik'/2} (1+\tanh (x))^{-ik/2} 
			{}_2F_1\left(\begin{matrix}
			\frac{1}{2}-\lambda-i\frac{k+k'}{2},\, \frac{1}{2}+\lambda-i\frac{k+k'}{2}
			\\[1ex]  1-i k \end{matrix}\, ;\frac{1+\tanh(x)}{2}\right)\,,
			\end{array}
\end{equation}
where $A$ and $B$ are arbitrary constants. The definition of the linearly independent solutions $\psi(x)$ and $\phi(x)$ of the Schr\"odinger equation is clear after \eqref{8}, where we assume that $1-i k$ is not a negative integer. For the hypergeometric function with parameters $a$, $b$, $c$ and indeterminate $z$, ${}_2F_1(a,b,c;z)$,  we have used the following notation
\begin{equation}\label{9}
_2F_1 \left( \begin{array}{c} a\; b \\ c \end{array}\,  ; z \right)\,.
\end{equation}	
Note that in \eqref{8}, we use $z= \frac{1+\tanh(x)}{2}$ as indeterminate. 

The objective of this discussion is finding the explicit form of the $S$-matrix, or scattering matrix, in order to find the structure of its analytic continuation to the complex plane. This may be performed either in the $k$ momentum representation or in the energy representation. We shall show that this objective can be realized. In fact, the $S$-matrix connects the asymptotic form of the solution as $x \longmapsto -\infty$ with the asymptotic form $x \longmapsto \infty$. The solvability of this problem comes from the knowledge of the asymptotic forms of the hypergeometric function.  Then, the asymptotic forms of \eqref{8} are:

\begin{itemize}

\item{On the limit as $x \longmapsto -\infty$. Note that $_2F_1(a,b,c;0)=1$. 
In this limit 
the asymptotic form $\Psi^-(x)$ of  $\Psi(x)$ given in \eqref{8} 
becomes

\begin{equation}\label{10}
\Psi^-(x) =  A^-\, e^{ikx} +  B^- \, e^{-ikx}\,,\qquad 
\ A^-= A\, 2^{i(k+k')/2}\,,\ B^-= B\, 2^{i(k+k')/2}
\end{equation}} 

\item{On the limit  $x \longmapsto \infty$ the situation looks a bit complicated. Nevertheless, we use here the following relation:

\begin{eqnarray}\label{11}
{}_2F_1(a,b,c;z)=\frac{\Gamma(c)\Gamma(c-a-b)}{\Gamma(c-a)\Gamma(c-b)}{}_2F_1(a,b,a+b+1-c;1-z)  \nonumber \\ [2ex] +\frac{\Gamma(c)\Gamma(a+b-c)}{\Gamma(a)\Gamma(b)} (1-z)^{c-a-b}{}_2F_1(c-a,c-b,1+c-a-b;1-z)\,,
\end{eqnarray}
which helps us to determine the asymptotic form $\Psi^+(x)$ of \eqref{8} as

\begin{eqnarray}\label{12}
\Psi^+(x) = A^+ \, e^{ik'x} + B^+ \, e^{-ik'x} \nonumber 
\\ [2ex] 
= \left( A\, 2^{i(k-k')/2} \; \frac{\Gamma\left(1+ik\right)\Gamma\left(ik'\right)}{\Gamma\left(\frac{1}{2}-\lambda+i\frac{k}{2}+i\frac{k'}{2}\right)\Gamma\left(\frac{1}{2}+\lambda+i\frac{k}{2}+i\frac{k'}{2}\right)}  \right. \nonumber 
\\ [2ex]
+ \left.  B\, 2^{i(k-k')/2}\; \frac{\Gamma\left(1+ik\right)\Gamma\left(-ik'\right)}{\Gamma\left(\frac{1}{2}-\lambda+i\frac{k}{2}-i\frac{k'}{2}\right)\Gamma\left(\frac{1}{2}+\lambda+i\frac{k}{2}-i\frac{k'}{2}\right)}  \right)\, e^{ik'x} \nonumber \\ [2ex] 
 + \left( A\, 2^{i(k+k')/2} \;\frac{\Gamma\left(1-ik\right)\Gamma\left(ik'\right)}{\Gamma\left(\frac{1}{2}-\lambda-i\frac{k}{2}+i\frac{k'}{2}\right)\Gamma\left(\frac{1}{2}+\lambda-i\frac{k}{2}+i\frac{k'}{2}\right)}  \right. \nonumber \\ [2ex]
 + \left. B\, 2^{i(k+k')/2} \;\frac{\Gamma\left(1-ik\right)\Gamma\left(-ik'\right)}{\Gamma\left(\frac{1}{2}-\lambda-i\frac{k}{2}-i\frac{k'}{2}\right)\Gamma\left(\frac{1}{2}-\lambda-i\frac{k}{2}-i\frac{k'}{2}\right)}\right) \, e^{-i k' x}\,.
\end{eqnarray}
}
\end{itemize}

From (\ref{10}) and (\ref{12}) we obtain the transition matrix, $\{T_{ij}\}$,

\begin{equation}\label{13}
\begin{pmatrix}
			A^+\\B^+
		\end{pmatrix}=\begin{pmatrix}
			T_{11}&T_{12}\\T_{21}&T_{22}\,
		\end{pmatrix}\begin{pmatrix}
			A^-\\B^-
		\end{pmatrix}
\end{equation}
which is given by

\begin{eqnarray}\label{14}
T(k,k') 
= {\Large  \begin{pmatrix}
				\frac{2^{-i k'}\Gamma\left(1+ik\right)\Gamma\left(ik'\right)}{\Gamma\left(\frac{1}{2}-\lambda+i\frac{k}{2}+i\frac{k'}{2}\right)\Gamma\left(\frac{1}{2}+\lambda+i\frac{k}{2}+i\frac{k'}{2}\right)}
				& \frac{2^{-i k'}\Gamma\left(1+ik\right)\Gamma\left(-ik'\right)}{\Gamma\left(\frac{1}{2}-\lambda+i\frac{k}{2}-i\frac{k'}{2}\right)\Gamma\left(\frac{1}{2}+\lambda+i\frac{k}{2}-i\frac{k'}{2}\right)}
				\\[2 ex]
				\frac{\Gamma\left(1-ik\right)\Gamma\left(ik'\right)}{\Gamma\left(\frac{1}{2}-\lambda-i\frac{k}{2}+i\frac{k'}{2}\right)\Gamma\left(\frac{1}{2}+\lambda-i\frac{k}{2}+i\frac{k'}{2}\right)}
				& \frac{\Gamma\left(1-ik\right)\Gamma\left(-ik'\right)}{\Gamma\left(\frac{1}{2}+\lambda-i\frac{k}{2}-i\frac{k'}{2}\right)\Gamma\left(\frac{1}{2}-\lambda-i\frac{k}{2}-i\frac{k'}{2}\right)} 
		\end{pmatrix}}\,.
\end{eqnarray}

Thus, we are in the position to give an expression for the scattering $S$-matrix,  relating ingoing $(A^-,B^+)$ and outgoing $(A^+,B^-)$ amplitudes, as
\begin{equation}\label{13b}
\begin{pmatrix}
			B^-\\A^+
		\end{pmatrix}=\begin{pmatrix}
			S_{11}&S_{12}\\S_{21}&S_{22}\,
		\end{pmatrix}\begin{pmatrix}
			A^-\\B^+
		\end{pmatrix}\,.
\end{equation}

Thus, the scattering matrix can be expressed in terms of the elements of the transition matrix in the following form

\begin{equation}\label{15}
S=\frac{1}{T_{22}}\begin{pmatrix}
			-T_{21}&1\\
			\det(T)& T_{12}	\end{pmatrix}\,.
\end{equation}

The interesting scattering features come from the poles of the scattering $S$-matrix \eqref{15}, which are the zeroes of $T_{22}$ that could be complex. Taking into account the explicit form of $T_{22}$ given in \eqref{14} and the properties of the Euler Gamma function, $\Gamma(z)$, we obtain two conditions that determine the zeroes of $T_{22}$. These are ($n=0,1,2,\dots$):

\begin{itemize}

\item{Condition 1: 

\begin{equation}\label{16}
\frac{1}{2}-\lambda-\frac{i}2(k+k')=-n \,.
\end{equation}
}

\item{Condition 2:

\begin{equation}\label{17}
\frac{1}{2}+\lambda-\frac{i}2(k+k')=-n\,.
\end{equation}}

\end{itemize}

These are two different conditions that will yield to different spectral properties, which we analyze next. In this analysis, we are to use the following relation between Jacobi polynomials and the hypergeometric functions (valid only when $a$ be a non-positive integer):

\begin{equation}\label{18}
_2F_1(a,b,c;z)=\frac{\Gamma (1-a) \Gamma (a-c+1) }{\Gamma (1-c)}P_{-a}^{(a+b-c,c-1)}(2 z-1)\,.
\end{equation}
Since for both conditions $k$ and $k'$ are pure imaginary numbers, we may use a special notation for their imaginary part. Thus, we define 

\[
\mu:=-ik',\qquad \nu:= -ik
\] as auxiliary parameters. Then, we study the consequences of each condition separately.

\medskip

{\it Condition 1}. It gives discrete values of the energy depending on $n$ as well as on $\lambda$, which are

\begin{equation}\label{19}
E^{\{1\}}_{\lambda,n}=\frac{k_{\lambda,n}^2+k_{\lambda,n}'^2}{2}=-\frac{\nu_{\lambda,n}^2+\mu_{\lambda,n}^2}{2}=-(\lambda-1/2-n)^2-\frac{\beta^2}{(\lambda-1/2-n)^2}\,.
\end{equation}
Note that we have labelled $k$ and $k'$, as well as $\nu$ and $\mu$ in terms of $n$ and $\lambda$. The first identity in \eqref{19} comes from \eqref{5} and \eqref{16}, the second after the definitions after \eqref{18} and the third is the consequence of \eqref{5} and \eqref{16} and some algebra. Now, after \eqref{16} and \eqref{5}, we have that

\begin{eqnarray}
\mu^{\{1\}}_{\lambda,n}=-i k'_{\lambda,n}=(\lambda-1/2-n)+\frac{\beta}{\lambda-1/2-n} \,, \label{20} \\ [2ex]  \nu^{\{1\}}_{\lambda,n}=-i k_{\lambda,n}=(\lambda-1/2-n)-\frac{\beta}{\lambda-1/2-n} \label{21}\,.
\end{eqnarray}

If we use the values $k_{\lambda,n}$ and $k'_{\lambda,n}$ given in \eqref{20} and \eqref{21} in the functions $\phi(x)$ and $\psi(x)$ of the general solution \eqref{8} as well as relation \eqref{18}, we obtain the following solutions indexed by $\lambda$ and $n$:

\begin{equation}\label{22}
\ps_{\lambda,n}(x)= e^{\frac{\beta}{\lambda-1/2-n}x}\operatorname{cosh}^{\lambda-1/2-n}(x) \, _2F_1\left(\begin{matrix}n+1,\,n-2 \lambda +1\\[1ex]  n-\lambda +\frac{ \beta }{\lambda-1/2-n}+\frac{3}{2}\end{matrix}; \frac{1+\tanh x}{2}\right)\,,
\end{equation} 
and

\begin{equation}\label{23}
\ph_{\lambda,n}(x)= e^{-\frac{\beta}{\lambda-1/2-n}x}\,\operatorname{sech}^{\lambda-1/2-n}(x)\,P_n^{(\mu^{\{1\}}_{\lambda,n},\nu^{\{1\}}_{\lambda,n})}\, (\tanh\, x)\,.
\end{equation}

In the search for scattering features, we should note that bound states may only appear when the potential has a well shape with a minimum, i.e., when $0<\beta<\left(\lambda^2-\frac{1}{4}\right)$. The well has a finite depth, with only a finite number of bound states \cite{MMN}. Then, some real poles of the scattering matrix expressed in terms of energies are expected to be eigenvalues of the Schr\"odinger equation with square integrable eigenfunctions. The solutions $\ps_{\lambda,n}(x)$ in \eqref{22} are never square integrable. For $\ph_{\lambda,n}(x)$ in \eqref{23}, we have three different possibilities. These are:

\begin{itemize}

\item{{\it Bound state poles}: 

Bound states correspond to simple poles of the scattering matrix on the energy representation, $S(E)$, with $E<0$. Each of these poles gives a solution of the Schr\"odinger equation with square integrable wave function. In the momentum representation, according to (\ref{2}), $k=\sqrt{E+2\beta}, k'=\sqrt{E-2\beta}$, these poles of the scattering matrix $S(k)$ are located on the positive side of the imaginary axis. 

In our terminology, bound state poles are characterized by the condition $\operatorname{Im}(k_{\lambda,n})>0,\,\operatorname{Im}(k'_{\lambda,n})>0$ or, equivalently, $\mu_{\lambda,n}>0,\,\nu_{\lambda,n}>0$ and are located at the negative energies  $\{E^{\{1\}}_{\lambda,n}\}_{n=0}^{n_{max}}$. Taking into account \eqref{20} and \eqref{21} and the above condition, the values of $n$ for which we have bound state poles are those satisfying simultaneously the conditions

\begin{equation}\label{24}
\mu^{\{1\}}_{\lambda,n}=(\lambda-1/2-n)+\frac{\beta}{\lambda-1/2-n} >0\,,\quad \nu^{\{1\}}_{\lambda,n}= (\lambda-1/2-n)-\frac{\beta}{\lambda-1/2-n}>0\,,
\end{equation}

which implies that $ 0 \le n < \lambda - 1/2 - \sqrt \beta$, so that $\lambda > 1/2+ \sqrt \beta$ is a necessary condition for the existence of bound states (here we always choose $\lambda>0$). In consequence,  $n_{max}=\lfloor \lambda-1/2-\sqrt{\beta}\rfloor$, where $\lfloor  k \rfloor$ denotes the entire part of the real number $k$. The corresponding wave functions, $\ps_{\lambda,n}$, are square integrable. 

In conclusion, the {\bf bound states} are labelled by $n$, under the conditions that:
\[
0 \le n < \lambda - 1/2 - \sqrt \beta,\qquad
0 \le n < n_{\rm max}, \qquad 
n_{\rm max}=\lfloor \lambda - 1/2 - \sqrt \beta\rfloor
\]

}

\item{{\it Redundant poles}:  

These poles of $S(E)$ are characterized by the condition $\operatorname{Im}(k_{\lambda,n})\,\cdot\,\operatorname{Im}(k'_{\lambda,n})<0\equiv \mu_{\lambda,n}\,\cdot\,\nu_{\lambda,n}<0$ and are located at the energies $\{E^{\{1\}}_{\lambda,n}\}_{n=n_{max}+1}^{n_r}$. A similar argument of that given in the previous paragraph shows that  $n_{r}=\lfloor \lambda-1/2+\sqrt{\beta}\rfloor$. Now, the eigenfunctions $\phi_{\lambda,n}$ are not square integrable, so that these poles do not give the energies of bound states. The existence of this kind of poles is due to the potential asymmetry and exists for values of $n$ on the interval $\lambda - 1/2 - \sqrt\beta \le n \le \lambda -1/2 + \sqrt \beta$.  Note that for $\beta=0$ this asymmetry disappears and in such a case, this type of poles does not exist. 
Wave function solutions of the Schr\"odinger equation with energies given by redundant poles are called {\it redundant states}. 

In summary, the {\bf redundant states} are labelled by $n$, under the conditions that:
\[
\lambda - 1/2 - \sqrt\beta \le n  \le \lambda -1/2 + \sqrt \beta,\qquad
n_{\rm max} \le n \leq n_{\rm r}, \qquad 
n_{\rm r}=\lfloor \lambda - 1/2 + \sqrt \beta\rfloor
\]
}

\item{{\it Anti-bound state poles}: 

These poles are located at values of the energy characterized by $\operatorname{Im}(k_{\lambda,n})<0,\,\operatorname{Im}(k'_{\lambda,n})<0\equiv \mu_{\lambda,n}<0,\,\nu_{\lambda,n}<0$ and are infinite in number: $\{E^{\{1\}}_{\lambda,n}\}_{n=n_r+1}^\infty$. In the momentum representation $k:=\sqrt{E}$, these poles of the scattering matrix $S(k)$ are located on the negative side of the imaginary axis. Antibound state poles appear for values of $n$ such that  $n > \lambda - 1/2 + \sqrt \beta$. 

Wave function solutions of the Schr\"odinger equation with energies given by antibound state poles are called {\it antibound states} or {\it virtual states}.  

Thus, the {\bf antibound states} are labelled by $n$,  under the conditions that:
\[
\lambda - 1/2 + \sqrt\beta < n,\qquad
n_{\rm r} < n , \qquad 
n_{\rm r}=\lfloor \lambda - 1/2 + \sqrt \beta\rfloor
\]

}

\item{{\it Resonance poles}: No resonance poles exists in this model, assuming $\lambda$ be real.}

\end{itemize}

\medskip

{\it Condition 2}. It comes obvious after \eqref{16} and \eqref{17} that 
Condition 2 can be obtained after a change $\lambda \longmapsto - \lambda$ 
from Condition 1. Let us write our results here just for completeness.  

Energy values:

\begin{equation}\label{25}
E^{\{2\}}_{\lambda,n}=E^{\{1\}}_{-\lambda,n}
=
-(\lambda+1/2+n)^2-\frac{\beta^2}{(\lambda+1/2+n)^2}\,.
\end{equation}
Thus, we have the following identity,
\begin{equation}\label{25b}
E^{\{2\}}_{\lambda-n-1,n}=E^{\{1\}}_{\lambda+n,n}
=
-(\lambda-1/2)^2-\frac{\beta^2}{(\lambda-1/2)^2}\,.
\end{equation}

Values of in and out momenta:

\begin{eqnarray}
\mu^{\{2\}}_{\lambda,n}=-i {k'}^{\{2\}}_{\lambda,n}=-(\lambda+1/2+n)-\frac{\beta}{\lambda+1/2+n}\,, \label{26} 
\\ [2ex]  
\nu^{\{2\}}_{\lambda,n}=-i k^{\{2\}}_{\lambda,n}=-(\lambda+1/2+n)+\frac{\beta}{\lambda+1/2+n}\,. \label{27}
\end{eqnarray}

Eigenfunctions for the values \eqref{25} of the energy:

\begin{equation}\label{28}
\Ph_{\lambda,n}(x)=\ph_{-\lambda,n}(x)= e^{-\frac{\beta}{\lambda+1/2+n}x}\,\operatorname{sech}^{-\lambda-1/2-n}(x)\,P_n^{(\mu^{\{2\}}_{\lambda,n},\nu^{\{2\}}_{\lambda,n})}\, (\tanh\, x)\,,
\end{equation}
and

\begin{equation}\label{29}
\psi^{(2)}_{\lambda,n}(x)=\psi^{(1)}_{-\lambda,n}(x)= e^{-\frac{\beta}{\lambda+1/2+n}x}\operatorname{cosh}^{-\lambda-1/2-n}(x) \, _2F_1\left(\begin{matrix}n+1,\,n+2 \lambda +1\\[2ex] n+\lambda -\frac{ \beta }{\lambda+1/2+n}+\frac{3}{2}\end{matrix}\;; \frac{1+\tanh x}{2}\right)\,.
\end{equation}

Under Condition 2, we may also find
redundant and anti-bound state poles for the scattering matrix $S(E)$ but no bound states.

{\bf Redundant poles} exists for those values of $n$ in the interval 
\begin{equation}
-\lambda - 1/2 - \sqrt\beta \le n \le -\lambda -1/2 + \sqrt \beta.
\end{equation} 

Finally, {\bf anti-bound state poles} appear for values of $n$ such that 
\begin{equation}
n > -\lambda - 1/2 + \sqrt \beta
\end{equation}
 and are infinite in number. We see that we have two lists of
redundant poles and anti-bound state poles, one for Condition 1 and the other for Condition 2. 

\begin{figure}
		\centering
		\includegraphics[width=0.45\textwidth]{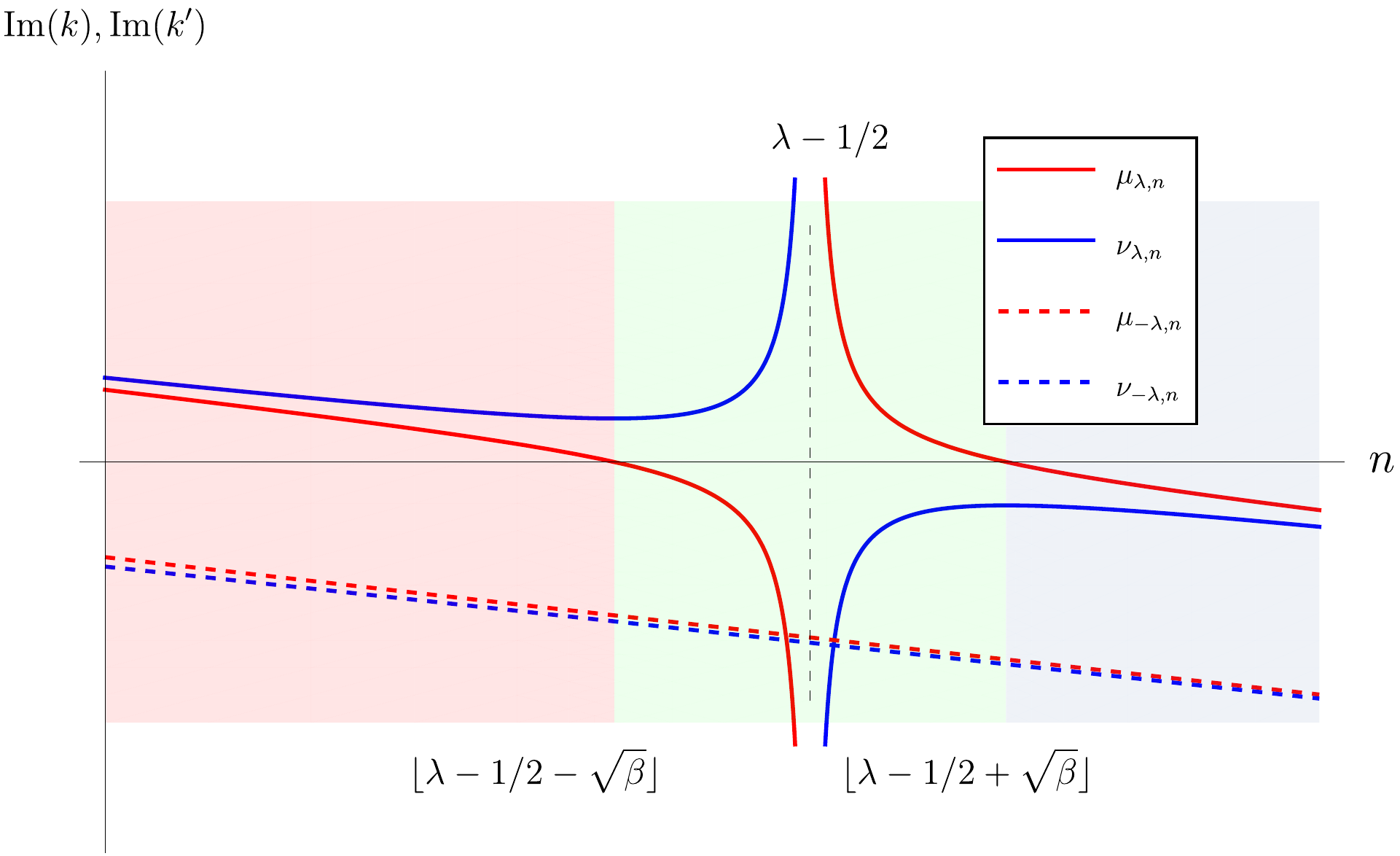}
		\includegraphics[width=0.4
	\textwidth]{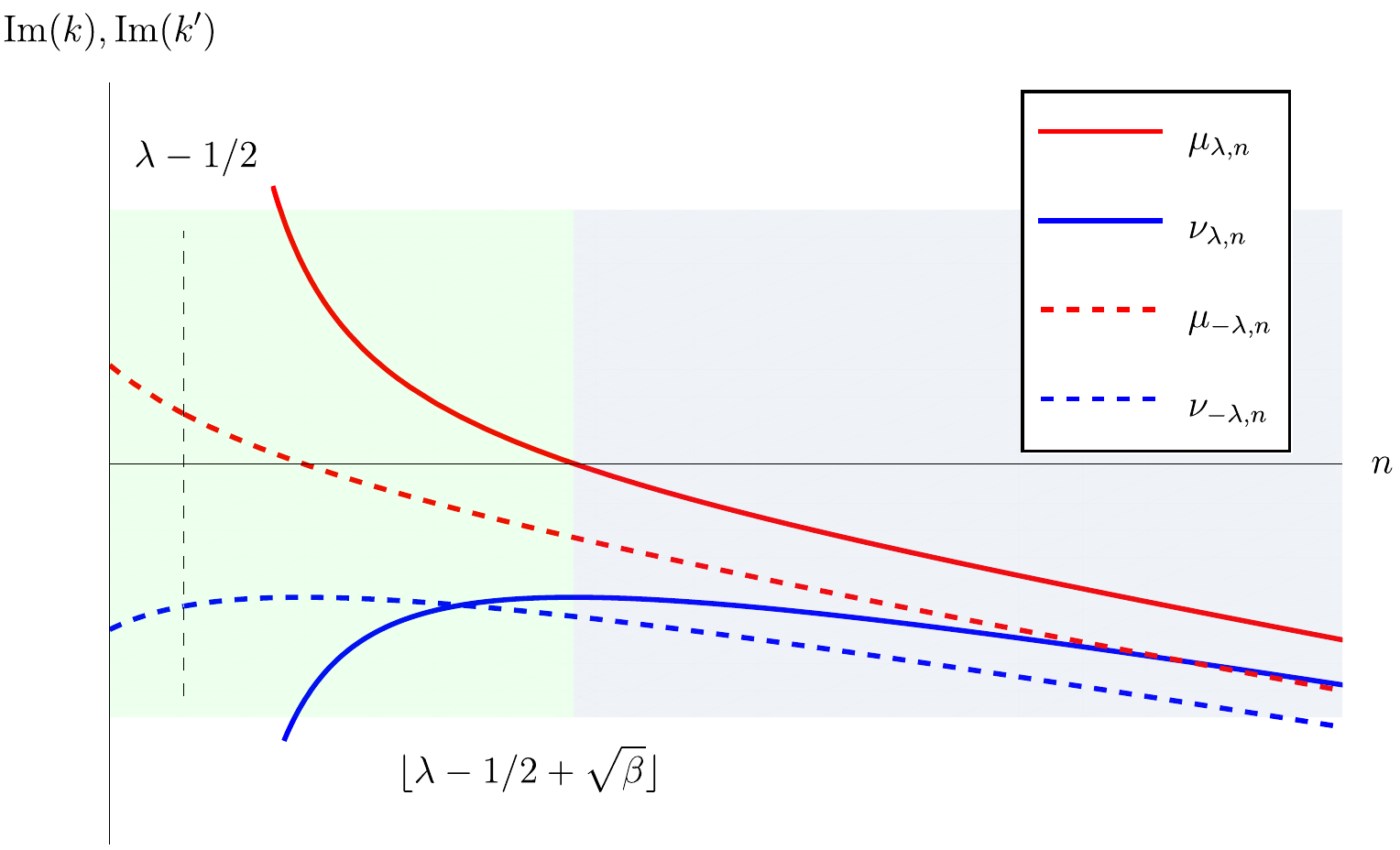}
		\caption{\small  Plot of $\nu^{\{1\}}_{\lambda,n}$ and $\mu^{\{1\}}_{\lambda,n}$: (left) $\lambda=4.1$, $\beta=1$ and (right)  $\lambda=1.1$, $\beta=10$, (Condition 1 in continuous  and 2 in dashing curves).  In the left plot: when both functions are positive ($0\leq n< \lambda-1/2-\sqrt\beta$), there are bound states. When both are negative, we are in the region of anti-bound states $n> \lambda-1/2 +\sqrt\beta$. Region in the middle, $\lambda-1/2 - \sqrt\beta < n <  \lambda-1/2 + \sqrt\beta$ corresponds to redundant poles, as one of the parameters is positive and the other negative. On the contrary, observe that values of $\nu_{-\lambda,n}$ and $\mu_{-\lambda,n}$ (Condition 2) are both negative for the chosen values of the parameters in the left plot, showing the presence of antibound states and no other features. In the right plot: for different values of the parameters $\lambda=1.1$ and $\beta=10$ there is no bound states for both Conditions 1 and 2, only redundant and antibound states exist.   }
			\label{fig:P1}
	\end{figure}

In Figure 2, we depict the behaviour of the imaginary parts of $k$ and $k'$ for both Conditions 1 and 2 and for given values of $\lambda>0$ and $\beta$. For the given value of the parameters, we observe 
that poles from Condition 1 may have bound, redundant and antibound states, while poles from Condition 2 may only have redundant or antibound states depending on the values of the parameters.

\subsubsection{Remark}

In a previous article \cite{GAKN}, we have studied the scattering matrix, $S(k)$, in momentum representation, for the one-dimensional Morse potential, where all poles of $S(k)$ are simple and lie on the imaginary axis. We have found series of poles including bound, anti-bound and redundant, for which their corresponding solutions are linked in each series by ladder operators. In such a case, the factorization of the Hamiltonian using ladder operators permits to relate states corresponding to neighbouring poles by these ladder operators, which are first order differential operators. A similar construction does not seem to be possible in here.

In \cite{GAKN}, it was shown that for the Hamiltonian with the one-dimensional Morse potential there are two first order differential ladder operators.
These ladder operators relate either bound-antibound series of states or two series of redundant states of the same Hamiltonian. For the Rosen--Morse II potential, this kind of first order ladder operators does not exist. Ladder operators could be only constructed as differential operators of $n$-th order, where $n$ is the order of the Jacobi polynomial that appear as a part of the eigenfunctions $\ph_{\lambda,n}(x)$ in \eqref{23} \cite{hussin19}. 

\sect{SUSY transformations}

In this second part of the paper, we are discussing three types of SUSY-partners of our original Hamiltonian using three different types of seed functions: ground state, redundant eigenfunctions and anti-bound eigenfunctions. We recall that the last two types of eigenfunctions are solutions of the Schr\"odinger equation with energies determined by a redundant pole or an anti-bound pole of the scattering matrix $S(E)$. Although these solutions are not square integrable,  they can be used to generate SUSY transformations. Let us discuss those three situations.

\subsection{SUSY with bound states and shape invariance potential}

Let us consider a nodeless  wave eigenfunction of the Rosen--Morse II Hamiltonian, i.e., a wave function without zeroes. Nodeless eigenfunctions for the Rosen--Morse II potentials have been classified in \cite{QUESNE}. In the set of bound states the only nodeless wavefunction  is that of the ground state, $\ph_{\lambda,0}(x)$, of $H_\lambda$ with energy $E^{\{1\}}_{\lambda,0}$.
As is customary in the theory of SUSY transformations, we define the superpotential derived from the ground state as

\begin{equation}\label{30}
W_\lambda(x)=-\frac{d}{dx}\ln \ph_{\lambda,0}=\frac{\beta}{\lambda-1/2}+(\lambda-1/2)\tanh x\,.
\end{equation}

The next step is the construction of the  first order factor differential operators $B_\lambda^{\mp}$ given by

\begin{equation}\label{31}
B_\lambda^{\mp}=\pm \frac{d}{dx}+W_\lambda (x)\,.
\end{equation}
From its construction, $B_\lambda^{-}\ph_{\lambda,0}=0$.
It happens that these  operators factorize the initial Hamiltonian $H_\lambda$ and its SUSY partner, which in this case is $H_{\lambda-1}$

\begin{equation}\label{factorizations}
H_\lambda= B_\lambda^{+}B_\lambda^{-}+ E^{\{1\}}_{\lambda,0},\qquad
H_{\lambda-1}=B_{\lambda}^{-}B_{\lambda}^{+}+ E^{\{1\}}_{\lambda,0}
\end{equation}

Replacing the expression on the superpotential 
$W_\lambda(x)$ in $H_\lambda$, we find the potential in terms of the superpotential,

\begin{equation}
H_\lambda=-\frac{d^2}{dx^2}+V_\lambda,\quad 
V_\lambda =  W^2_\lambda(x) - 
W'_\lambda(x)-E^{\{1\}}_{\lambda,0}
\end{equation}

and from (\ref{factorizations}), the potential $V_{\lambda-1}$ is given by 

\begin{equation}\label{vlambda}
V_{\lambda-1} = V_{\lambda}+ 2 
W'_\lambda(x) = V_\lambda -2 \frac{d^2}{dx^2}\ln \ph_{\lambda,0}
\end{equation}

From these factorizations the following intertwining relations are satisfied for any two consecutive Hamiltonians, 
$\{ H_{\lambda}, H_{\lambda-1}\}$:

\begin{equation}\label{33}
B^-_\lambda H_{\lambda}= H_{\lambda-1} B^-_\lambda,\qquad H_{\lambda} B^+_\lambda= B^+_\lambda H_{\lambda-1}\,,
\end{equation}

This is the property of {\it shape invariance} of the  Rosen--Morse II hierarchy $H_{\lambda+n}$, $n\in \mathbb{Z}$, carried out by the operators $B^\pm_{\lambda+n}$. This means that the ordinary SUSY transformations (by ordinary, we mean those SUSY transformations that make use of the ground state as seed function, which is the case here), gives rise to the original potential with a shift in the parameters, in particular, $\lambda$. The shift operator $B^-_\lambda$, changes $H_{\lambda}$ into $H_{\lambda-1}$ having one bound state less. This is a consequence to the fact that the resulting potential is less deep. One obtains the same result after successive applications of the SUSY transform, so that the potential well becomes more and more shallow. Consequently, after a finite number of transformations and on, the resulting potential cannot have bound states. We depict this effect in Figure 3.

An immediate consequence of (\ref{33}) is that 
$B^-_\lambda$ and $B^+_\lambda$ connect the eigenfunctions of  $H_\lambda$ and $H_{\lambda-1}$ as discussed in the formulas just below, where we denote by $\mathcal H_\lambda$ to the space spanned by all the eigenfunctions $H_\lambda$ with fixed $\lambda$.
After the properties of the Jacobi polynomials and the hypergeometric functions, in particular their behaviour with respect to differentiation \cite{OLVER}, we obtain the following action of the intertwining operators $B_\lambda^\pm$ on the outgoing wave functions resulting from the Conditions 1 and 2:

\begin{equation}\label{bimages1}
B_\lambda^-: 
\left\{
\begin{array}{ll}
{\cal H}_\lambda\ \to & {\cal H}_{\lambda-1}
\\[2.5ex]
\ph_{\lambda,n} \ \propto &\ph_{\lambda-1,n- 1}
\\[2.ex]
\ps_{\lambda,n} \ \propto &\ps_{\lambda-1,n- 1}
\end{array}
\right.
\quad 
B_\lambda^+: 
\left\{
\begin{array}{ll}
{\cal H}_{\lambda-1}\ \to & {\cal H}_\lambda
\\[2.5ex]
\ph_{\lambda-1,n-1} \ \propto &\ph_{\lambda,n}
\\[2.ex]
\ps_{\lambda-1,n-1} \ \propto &\ps_{\lambda,n}
\end{array}
\right.\quad n=1,2,\dots
\end{equation}
\medskip

\begin{equation}\label{bimages2}
B_\lambda^-: 
\left\{
\begin{array}{ll}
{\cal H}_\lambda\ \to & {\cal H}_{\lambda-1}
\\[2.5ex]
\Ph_{\lambda,n} \ \propto &\Ph_{\lambda-1,n+ 1}
\\[2.ex]
\Ps_{\lambda,n} \ \propto &\Ps_{\lambda-1,n+ 1}
\end{array}
\right.
\quad
B_\lambda^+: 
\left\{
\begin{array}{ll}
{\cal H}_{\lambda-1}\ \to & {\cal H}_\lambda
\\[2.5ex]
\Ph_{\lambda-1,n+1} \ \propto &-\Ph_{\lambda,n}
\\[2.ex]
\Ps_{\lambda-1,n+1} \ \propto &-\Ps_{\lambda,n}
\end{array}
\right. \qquad n=0,1,2,\dots
\end{equation}

where $n\neq 0$ in the action of $B_\lambda^-$ and $n\neq -1$
for $B_\lambda^+$. For these particular cases (or, in other words, for ground states transformations),
we have the following relations:

\begin{equation}\label{38}
B_\lambda^- \ph_{\lambda,0}= 0\,,\qquad B_\lambda^+ \Ph_{\lambda-1,0}= 0\,
\end{equation}
and
\begin{equation}\label{38b}
B_\lambda^- \ps_{\lambda,0}\propto \Ph_{\lambda-1,0}\,,\qquad B_\lambda^+ \Ps_{\lambda-1,0}\propto \ph_{\lambda,0}\,.
\end{equation}

Then, we have for each $\lambda>1/2+\sqrt{\beta}$ (in case $\lambda<1/2+\sqrt{\beta}$, we obtain redundant states) the following sequence of bound states  obtained by application of $B_{\lambda}^+$:

\begin{equation}
\ph_{\lambda,0}\ \xrightarrow{B^+_{\lambda+1}}\  \ph_{\lambda+1,1}\ \xrightarrow{B^+_{\lambda+2}}\ph_{\lambda+2,2} \dots
\ \xrightarrow{B^+_{\lambda+n}}\ \ph_{\lambda+n,n} \dots
\end{equation}

 and by application of $B_{\lambda}^-$ we find  
 2-antibound states:

\begin{equation}
\Ph_{\lambda-1,0}\ \xrightarrow{B^-_{\lambda-1}}\ \Ph_{\lambda-2,1}\ \xrightarrow{B^-_{\lambda-2}}\ \Ph_{\lambda-3,2} \dots
\xrightarrow{B^-_{\lambda-n}}\ \ \Ph_{\lambda-n-1,n} \dots\,.
\end{equation}

Both sequences have the same energy: 
$E^{\{1\}}_{\lambda+n,n}=E^{\{2\}}_{\lambda-n-1,n}$, $n=1,2,\dots$ as shown in (\ref{25b}).

One consequence of these ``ordinary'' SUSY transformations of the pair $H_\lambda, H_{\lambda-1}$ based on the ground states is that the action of 
$B_{\lambda}^-$ eliminates one  bound state $\ph_{\lambda,0}$ of energy $E^{\{1\}}_{\lambda,0}$ of $H_\lambda$, so that this point is not in  the spectrum of $H_{\lambda-1}$.  On the other hand, $B_{\lambda}^+$ eliminates one antibound state $\Ph_{\lambda-1,0}$ of $H_{\lambda-1}$ with the same energy which
is not in $H_\lambda$. In other words,
this transformation from $H_\lambda$ to $H_{\lambda-1}$
interchanges one bound state (in the initial Hamiltonian but not in its partner) by one 2-antibound virtual state (present in the partner, but not in the initial one) of the same energy. The rest of poles and states remain unaltered.

\begin{figure}[h!]
	\hskip-1.cm\includegraphics[width=1.1\linewidth]{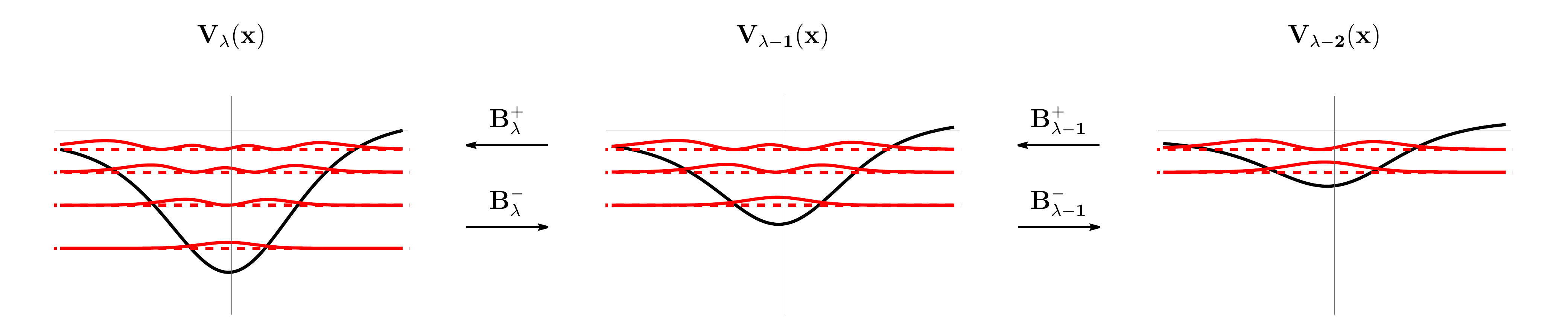}
	\caption{Rosen--Morse II potential, $V_\lambda(x)$ with $\lambda=5.4$ and $\beta=1$. After two successive SUSY transformations (\ref{vlambda}), we see that one looses one ground state after the first one and two after the second, so that only two ground states remain. The potential $V_{\lambda-4}(x)$ will  have no bound states.}
	\label{fig:SUSYn}
\end{figure}

Speaking about successive SUSY transformations, the result of $\ell$ transformations on the original state wave function $\ph_{\lambda,n}$ can be obtained either by multiplying successive intertwining operators or as the Wronskian of the lost levels as:

\begin{equation}\label{40}
B^-_{\lambda+1-\ell}\cdots	B_{\lambda-1}^-	B_\lambda^- \ph_{\lambda,n}=\frac{\mathcal{W}\left(\{\ph_{\lambda,j}\}_{j=0}^{\ell-1}\cup \{\ph_{\lambda,n}\}\right)}{\mathcal{W}\left(\{\ph_{\lambda,j}\}_{j=0}^{\ell-1}\right)}\propto \ph_{\lambda-\ell,n-\ell}\,.
\end{equation}

The resulting potential after $\ell$ transformation becomes 

\begin{equation}\label{41}
V_{\lambda-\ell}=V_\lambda-2\frac{d^2}{dx^2}\ln \mathcal{W}\left(\{\ph_{\lambda,j}\}_{j=0}^{\ell-1}\right)
\end{equation}

Once we have analyzed the action of successive SUSY transformations using the original potential ground state as seed function, we wish to use another type of seed functions, assuming them nodeless, such as  the wave function of a redundant state pole or an anti-bound state. We shall do it in here next. 

First of all, we give an overview on the possibilities of finding nodeless solutions to the Schr\"odinger equation with eigenvalues either redundant or anti-bound poles of the scattering matrix $S(E)$.
\subsection{Classification of nodeless solutions.}

In order to construct 1-SUSY transforms such that the new potentials have more levels than the original one, we need nodeless seed functions, i.e., functions without zeroes on the real axis. Solutions of the Schr\"odinger equation with energies given by redundant poles (redundant states) or anti-bound poles (anti-bound or virtual states) may be suitable to act as seed functions. We need criteria to know when these functions are nodeless. To this end, following the classification by Quesne in \cite{QUESNE} of the nodeless solutions of the Rosen--Morse II Schr\"odinger equation, there are three types of these nodeless solutions:

Type $I$ nodeless solutions. Those solutions $\phi^I_{\lambda,m}(x)$ with 

\begin{equation}\label{44}
m\in \mathbb{N},\quad\lambda-1/2>m,\quad (\lambda-1/2)(\lambda-1/2-m)<\beta<(\lambda-1/2)^2\,.
\end{equation}

Type $II$ nodeless solutions, $\phi^{II}_{\lambda,m}(x)$, characterized by

\begin{equation}\label{45}
m\in \mathbb{N},\quad m/2<\lambda-1/2<m,\quad -(\lambda-1/2)(\lambda-1/2-m)<\beta<(\lambda-1/2)^2\,.
\end{equation}

Type $III$ nodeless solutions, $\phi^{III}_{\lambda,m}(x)$, characterized by

\begin{equation}\label{46}
m/2\in \mathbb{N},\quad\lambda>1/2- \sqrt\beta,\quad 0<\beta<(\lambda-1/2)^2\,.
\end{equation}

In order to obtain nodeless solutions of type $I$ or $II$, we need to have a large value of the parameter $\beta$. Solutions for redundant poles with the Condition 1 are of either one of these types, while those of type $III$ correspond to anti-bound states with Condition 2. In the latter case, nodeless solutions appear no matter the value of $\beta>0$. 
\subsection{SUSY transformation with a seed function given by a redundant state.}

Redundant poles appear for large values of $\beta$, so that we choose high $\beta$ in this discussion. However, large values of $\beta$ may result in the existence of very few or none bound states. Let us give a specific example. If we choose $\lambda=5.4$ and $\beta=6$, the Rosen--Morse II Hamiltonian shows three bound states plus one nodeless redundant state. This redundant state is $\ph_{\lambda,m}(x)$ with $\lambda=5.4$ and $m=4$ and is of type $I$, according to the classification given in the previous subsection. In the present case ($\lambda=5.4, m=4,\beta=6$), we have the following pair of supersymmetric Hamiltonians:

\begin{eqnarray}\label{47}
{\widetilde H}=H_{\lambda}-2\frac{d^2}{dx^2}\ln \ph_{\lambda,\,m}(x)\,,
\end{eqnarray}

and therefore

\begin{eqnarray}\label{47b}
{\widetilde V}=V_{\lambda}-2\frac{d^2}{dx^2}\ln \ph_{\lambda,\,m}(x)\,,
\end{eqnarray}

where we have used the symbol ``tilde'' in order to underline the difference between the ``ordinary'' transformation where one makes use of the ground state from this one. 
The SUSY operator connecting both Hamiltonians is defined as in (\ref{31}) by
\begin{equation}\label{31b}
\tilde B_\lambda^{\mp}=\pm \frac{d}{dx}+W_\lambda (x) =\pm \frac{d}{dx} -\frac{d}{dx}\ln \ph_{\lambda\,, m}\,.
\end{equation}
Then, the operator $B_\lambda^{-}$ annihilates $\ph_{\lambda\,, m}$ of 
$H_\lambda$, while
$B_\lambda^{+}$ will annihilate $ \tilde \phi_{\lambda\,, m}$ of $\widetilde H$:
\begin{equation}\label{31bb}
B_\lambda^{-}\ph_{\lambda\,, m}=0,\qquad 
B_\lambda^{+}\tilde \phi_{\lambda\,, m}=0,\qquad \tilde \phi_{\lambda\,, m}=\frac1{\ph_{\lambda\,, m}} \, .
\end{equation}
In any case, the energy of the seed function can be shifted so that it be located at zero by substracting $E^{\{1\}}_{\lambda,\, m}$ in \eqref{47}. In Figure 4, we show how the first order transformation affects to Hamiltonians and bound states.

\begin{figure}[h!]
	\hskip-0.8cm \includegraphics[width=1.\linewidth]{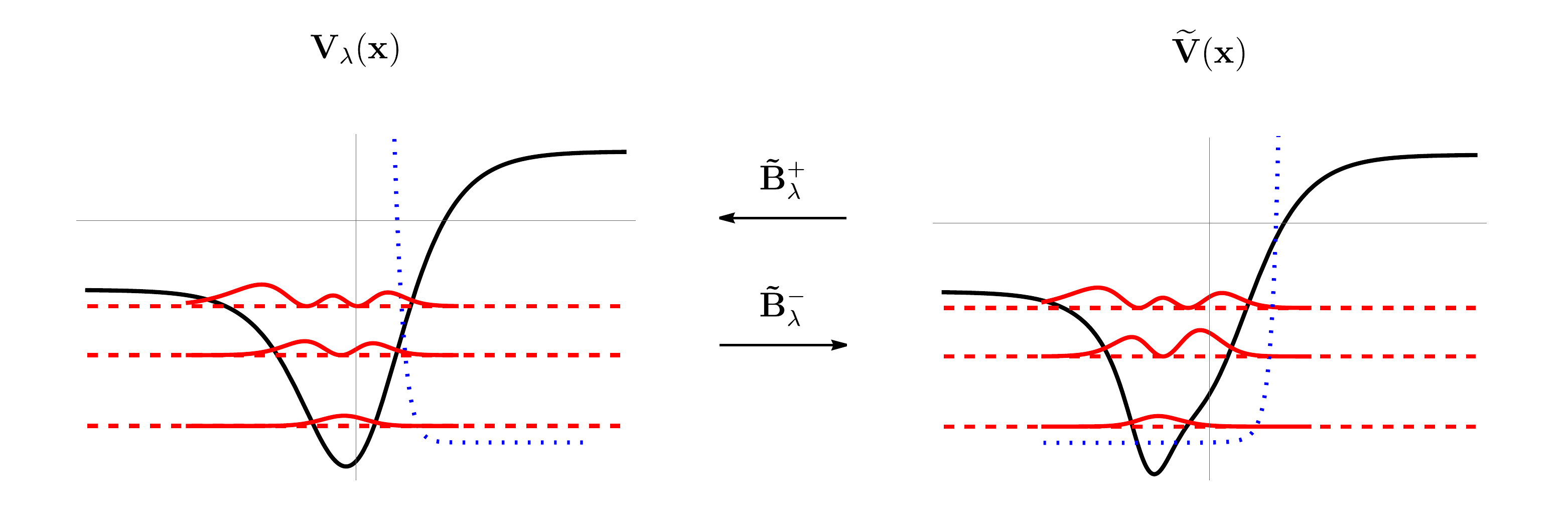}
	\caption{At the left, a Rosen--Morse II potential with $\lambda=5.3$ and $\beta=4$. At the right the result of a first order SUSY transformation (\ref{47b}) on the potential on the left that uses a nodeless redundant solution of type $I$ on dotted blue. Note that the transformation (\ref{31b}) preserves the number of bound states, since the new one results on a wave function which is not square integrable. This is depicted in dotted blue at the right.  }
	\label{fig:SUSY4a4}
\end{figure}

Note that this transformation does not produce a new bound state. In this example, three bound states for ${H}_{\lambda}$ result on three bound states for $\widetilde{H}_{\lambda}$ . Instead to a new bound state, we have a wave function which is not square integrable. It is solution of the Schr\"odinger equation with energy given by the energy of the redundant pole.  This wave function is depicted on dotted blue in the right hand side of Figure 4. This resulting wave function corresponds again to a redundant pole.

\subsection{SUSY transformation with a seed function given by a anti-bound state.}

Now, the point of departure is a Rosen--Morse II potential with parameters $\lambda=2.4$ and $\beta=1$. This Hamiltonian has only one bound state. A nodeless solution of the Schr\"odinger equation of type $III$ gives the wave function for an anti-bound state. We propose as a seed function $\Ph_{\lambda,m}(x)$ with $\lambda= 2.4$ 
and $m=2$, which fulfills all these conditions. This function is given by a Jacobi polynomial of order two, which is nodeless. In this case ($\lambda=2.4, \beta=1,m=2$), we have the following pair of Hamiltonians related by first order transformation:
\begin{eqnarray}\label{48}
\widetilde{V}=
{V}_{\lambda}-2\frac{d^2}{dx^2}\ln \Ph_{\lambda,\,m}(x)
\end{eqnarray}
\begin{equation}\label{31c}
\tilde B_\lambda^{\mp}=\pm \frac{d}{dx} -\frac{d}{dx}\ln \Ph_{\lambda\,, m}\,.
\end{equation}

This transformation is depicted in Figure 5. In this case, the new potential after the SUSY transformation acquires a new bound state.

\begin{figure}[t]
		\centering
		\includegraphics[width=1.\linewidth]{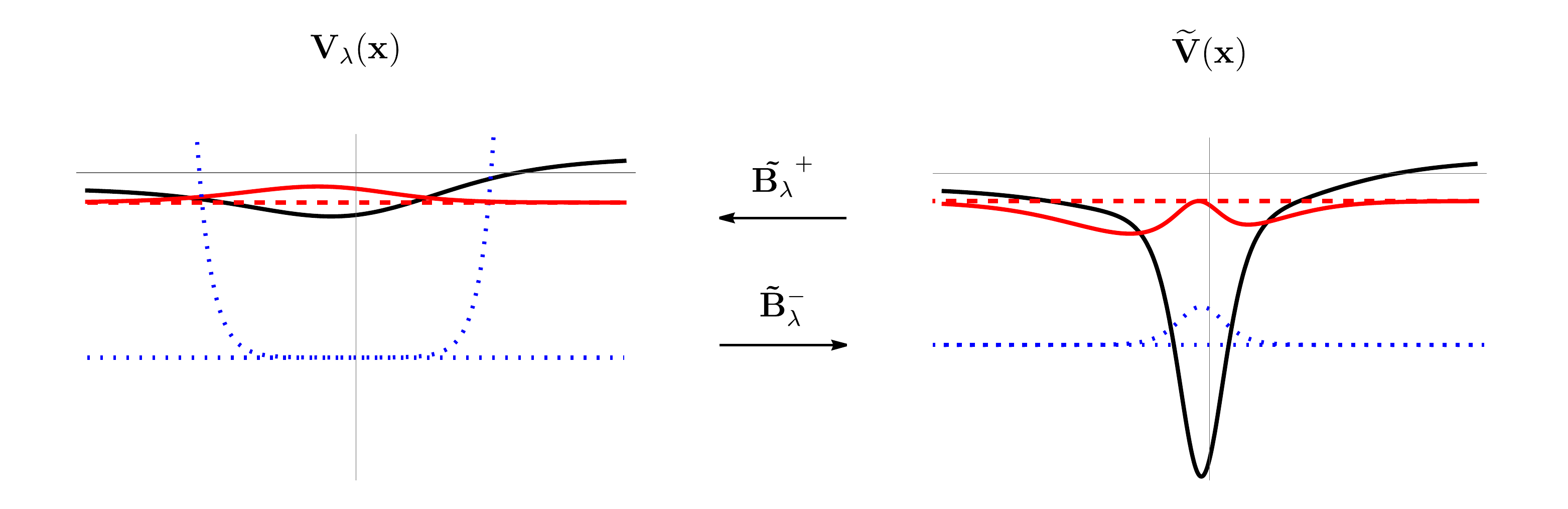}
		\caption{At the left, Rosen--Morse II potential of (\ref{48}) with  $\lambda=2.4$ and $\beta=1$ bearing an only bound state (in red). In dotted blue the anti-bound state wave function. Its SUSY transformation appears on the right hand side. It has two bound states, one in red and the new one in dotted blue. The SUSY operators $\tilde B_\lambda^{\mp}$ correspond to (\ref{31c}).}
		\label{fig:SUSY1a2}
\end{figure}

Note that whenever the transformation is produced by a real anti-bound wave function, then one new bound state arises after the transformation. This is a consequence of formula \eqref{31bb}, where we write the explicit form of the first ground state after the transformation. The function in the denominator of \eqref{31bb} is the wave function for the anti-bound state, which diverges exponentially on both sides. Therefore, its inverse is square integrable. This situation does not arise after a redundant state, which is rapidly increasing in one side, while is bounded on the other side, see Figure 4. 

It is interesting to show that there are at least two ways to arrive to the Hamiltonian $\widetilde H$ by means of SUSY transformations. One has been just described above. The point of departure for another one is the Rosen--Morse potential $V_{\lambda}$ depicted in Figure 3, left. The parameters here were $\lambda=5.4$, $\beta=1$. This Hamiltonian has four bound states, three redundant states and an infinite number of anti-bound states. Starting from its ground state, we perform a second order SUSY transformation, resulting in a Hamiltonian with only two bound states as the first and second  excited states in the original Hamiltonian have been erased.  These two Hamiltonians under consideration are (see \eqref{41}):

\begin{eqnarray}\label{49b}
\widetilde{\widetilde{V}}=V_\lambda-2\frac{d^2}{dx^2}\ln \mathcal{W}\left(\{\ph_{\lambda,1},\ph_{\lambda,2}\}\right)\,.
\end{eqnarray}

We first construct the second order operator $\tilde B_\lambda^{-}$, taking into account that the intertwining is given in terms of Wronskians

\begin{eqnarray}\label{49c}
\widetilde{\widetilde{\psi}}=\tilde B_\lambda^{-}\psi:=  \frac{\mathcal{W}\left(\{\ph_{\lambda,1},\ph_{\lambda,2},\psi\}\right)}{\mathcal{W}\left(\{\ph_{\lambda,1},\ph_{\lambda,2}\right)}\,.
\end{eqnarray}

The operator $\tilde B_\lambda^{+}$ can be defined as the Hermitian conjugate of  
$\tilde B_\lambda^{-}$.

This second order transformation is depicted in Figure 6. A comparison between Figures 5 and 6 show that $\tilde V $ in \eqref{48} and $\widetilde{\widetilde{V}}$ in \eqref{49b} are indeed the same Hamiltonian. This is a particular case of a more general situation as shown on the next Section.

\begin{figure}[h!]
	\centering
	\includegraphics[width=1.\linewidth]{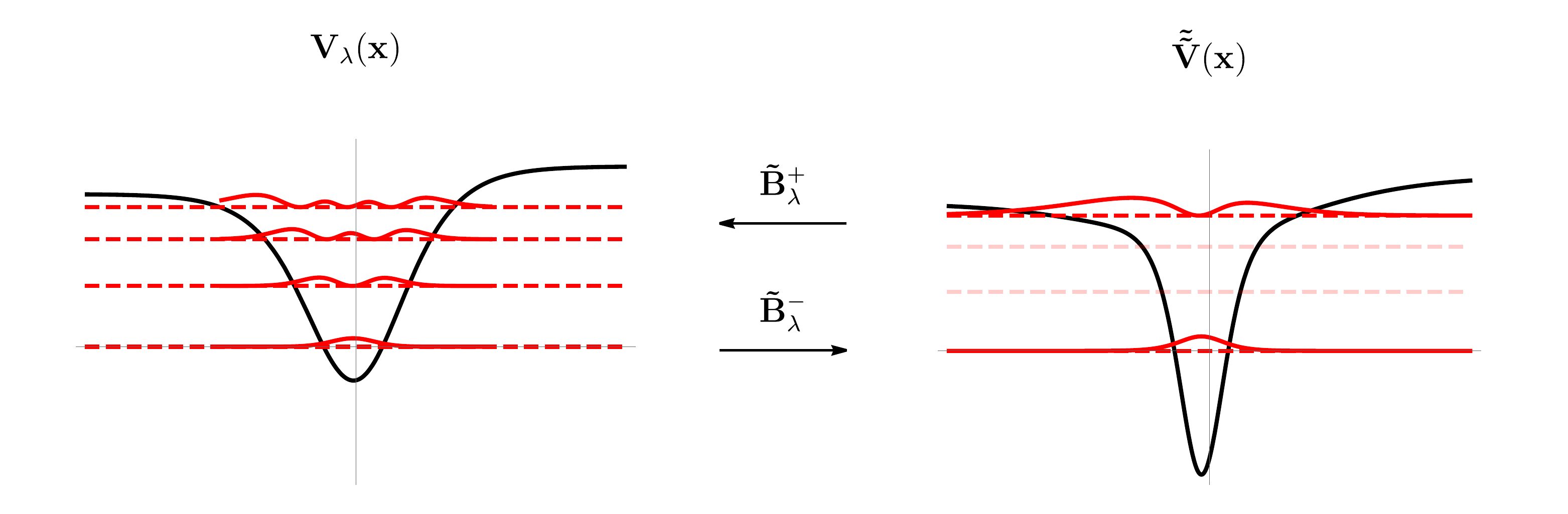}
	\caption{At the left a Rosen--Morse II potential with $\lambda=5.4$ and $\beta=1$. After a second order SUSY transformation (\ref{49b}), we arrive to the situation depicted on the right hand side. Observe that it is the same result than the obtained after the transformation in Figure 5.  This second order transformation erases two levels from the original Hamiltonian.}
	\label{fig:SUSY4a2}
\end{figure}

\sect{On the equivalence between different SUSY transformations of the Rosen--Morse II potential for $N$ odd.}

The situation described in the previous subsection is rather general as we intend to show along the present Section. We start by a Hamiltonian $H_\lambda$ with one bound state,
$\ee_{\lambda,0}$. Then, by means of a 2-antibound state $\Ph_{\lambda,N-1}$ the corresponding SUSY transformed Hamiltonian $\tilde H^{(1)}_\lambda$ will have two bound states with energies
$\ee_{\lambda,0}$ and $\Ee_{\lambda,N-1} = \ee_{\lambda+N,0}$ (where $^{(1)}$ corresponds to the order of the SUSY transformation). Therefore, 
$\tilde H^{(1)}_\lambda$ has two bound states with energies $\ee_{\lambda+N,0}$
and $\ee_{\lambda,0}=\ee_{\lambda+N,N}$. 

Next, we consider the Hamiltonian 
$H_{\lambda+N}$. It is clear that $H_{\lambda+N}$ can be obtained from $H_{\lambda}$ by
means of the application of $N$ sucesive operators $B^+$, therefore, it has
$N+1$ bound states with energies: $\ee_{\lambda+N,0}, \ee_{\lambda+N,1}, \dots
\ee_{\lambda+N,N}$. Therefore, if we eliminate the $N-1$ intermediate levels by a 
$(N{-}1)$-order SUSY transformation we will get a Hamiltonian $\tilde H_{\lambda+N}^{(N-1)}$ with
only two levels: $\ee_{\lambda+N,0}$ and $\ee_{\lambda+N,N}=\ee_{\lambda,0}$. Which is the same spectrum of $\tilde H^{(1)}_\lambda$. In fact,  both Hamiltonians are identical:
 
\begin{equation}
    \tilde H^{(1)}_\lambda = \tilde H_{\lambda+N}^{(N-1)}
    \end{equation}

This is proved in the rest of this section.

Let us consider a one dimensional Rosen--Morse II Hamiltonian $H_\alpha:= -d^2/dx^2 + V_\alpha(x)$, and $\alpha\in(1/2+\sqrt{\beta},3/2+\sqrt{\beta})$. These Hamiltonians have a unique bound state. However, if we replace $\alpha$ by $\alpha + N$, where $N$ is a natural number, $N=1,2,\dots$, then the Hamiltonian has $N+1$ bound states.  

To begin with, let us consider $H_\alpha$ with $\alpha\in(1/2+\sqrt{\beta},3/2+\sqrt{\beta})$. Take an anti-bound state, $\Ph_{\alpha,N-1}(x)$ of type $III$. Its SUSY transform of first order has two bound states. The situation is as follows: If we denote by $E^{\{2\}}_{\alpha,N-1}$ the energy corresponding to this anti-bound state, and given by its corresponding pole on $S(E)$, the original Hamiltonian and its first order partner are, respectively,

\begin{eqnarray}\label{50}
 \widetilde{H}^{(1)}_\alpha=H_\alpha-2\frac{d^2}{dx^2} \ln \Ph_{\alpha,N-1}
\end{eqnarray}

The Hamiltonian $ \widetilde{H}^{(1)}_\alpha$ has, consequently, two bound states. The energy and the wave function of the ground state are, respectively, 

\begin{equation}\label{51}
\widetilde{E}^{(1)}_{\alpha,0}
\,, \qquad \widetilde{\phi}_{\alpha,0}=\frac{\mathcal{W}(\Ph_{\alpha,N-1},\Ps_{\alpha,N-1})}{\Ph_{\alpha,N-1}}=\frac{1}{\Ph_{\alpha,N-1}}\,,
\end{equation}

while the energy and the wave function of the excited state are given, respectively, by

\begin{equation}\label{52}
\widetilde{E}^{(1)}_{\alpha,1}=E^{\{1\}}_{\alpha,0}\,, \qquad \widetilde{\phi}_{\alpha,1}=\frac{\mathcal{W}(\Ph_{\alpha,N-1},\ph_{\alpha,0})}{\Ph_{\alpha,N-1}}=\ph_{\alpha,0}\frac{\Ph_{\alpha-1,N}}{\Ph_{\alpha,N-1}}\,.
\end{equation}

Next, we consider the Hamiltonian $H_{\alpha + N}$ and  the SUSY transformation of order $N-1$, which gives as final result:

\begin{equation}\label{54}
\widetilde{H}^{(N-1)}_{\alpha+N}=H_{\alpha+N}
-2\frac{d^2}{dx^2} \ln \mathcal{W}\left(\left\{\ph_{\alpha+N,j}\right\}_{j=1}^{N-1}\right)\,,
\end{equation}

where the Wronskian does not include the functions $\phi_{\alpha,0}(x)$ and  $\phi_{\alpha,N-1}(x)$. After this transformation, two bound states are left from the $N+1$ bound states of the original Hamiltonian. The energy and wave function of the ground state are, respectively,

\begin{equation}\label{55}
\widetilde{E}^{(N-1)}_{\alpha,1}
\,, \qquad \widetilde{\phi}^{(N-1)}_{\alpha,0}=\frac{\mathcal{W}\left(\left\{\ph_{\alpha+N,j}\right\}_{j=1}^{N-1} \cup \{\ph_{\alpha+N,0}\}\right)}{\mathcal{W}\left(\left\{\ph_{\alpha+N,j}\right\}_{j=1}^{N-1}\right)}\,,
\end{equation}

while the same data corresponding to the excited state are

\begin{equation}\label{56}
\widetilde{E}^{(N-1)}_{\alpha,1}=E^{\{1\}}_{\alpha,0}
\,, \quad \widetilde{\phi}^{(N-1)}_{\alpha,1}=\frac{\mathcal{W}\left(\left\{\ph_{\alpha+N,j}\right\}_{j=1}^{N-1} \cup \{\ph_{\alpha+N,N}\}\right)}{\mathcal{W}\left(\left\{\ph_{\alpha+N,j}\right\}_{j=1}^{N-1}\right)}\,.
\end{equation}

The objective is to show that

\begin{equation}\label{57}
\widetilde{H}^{(N-1)}_{N +\alpha} = \widetilde H_\alpha^{(1)}\,.
\end{equation}

With this objective in mind, let us make the following sequence of operations:

\begin{eqnarray}\label{58}
\widetilde{H}^{(N-1)}_{\alpha+N}=H_{\alpha+N}-2\frac{d^2}{dx^2} \ln \mathcal{W}\left(\left\{\ph_{\alpha+N,j}\right\}_{j=1}^{N-1}\right) \nonumber 
\\ [2ex] 
=-\frac{d^2}{dx^2}+V_{\alpha+N}-2\frac{d^2}{dx^2} \ln \mathcal{W}\left(\left\{\ph_{\alpha+N,j}\right\}_{j=1}^{N-1}\right)  \nonumber 
\\ [2ex] 
=H_{\alpha}	
-2\frac{d^2}{dx^2} \ln \left(\frac{\mathcal{W}\left(\left\{\ph_{\alpha+N,j}\right\}_{j=0}^{N-1}\cup\{\ps_{\alpha+N,0}\}\right)}{\mathcal{W}\left(\left\{\ph_{\alpha+N,j}\right\}_{j=0}^{N-1}\right)}\right)\,.
\end{eqnarray}

Here, the function $\ps_{\alpha+N,0}$, the non square integrable second solution of the Schr\"odinger equation with energy $E^{\{2\}}_{\alpha,N-1}$, is being transformed through a  $(N-1)$-th order SUSY transformation:

\begin{equation}\label{59}
\frac{\mathcal{W}\left(\left\{\ph_{\alpha+N,j}\right\}_{j=0}^{N-1}\cup\{\ps_{\alpha+N,0}\}\right)}{\mathcal{W}\left(\left\{\ph_{\alpha+N,j}\right\}_{j=0}^{N-1}\right)}=B^-_{\alpha}\cdots B^-_{\alpha+N}\ps_{\alpha+N,0}\,.
\end{equation}

The action of the annihilation operator on $\psi_{\alpha+N,0}$ is given by

\begin{equation}\label{60}
B^{-}_{\alpha+N}\ps_{\alpha+N,0}\propto\Ph_{\alpha+N-1,0}\,.
\end{equation}

Then, recalling \eqref{bimages2}, we finally obtain

\begin{eqnarray}\label{61}
B^{-}_{\alpha}\cdots B^{-}_{\alpha+N}\ps_{\alpha+N,0}=B^{-}_{\alpha}\cdots B^{-}_{\alpha+N-1}\Ph_{\alpha+N-1,0} \propto (-1)^{N-1}\Ph_{\alpha,N-1}\,.
\end{eqnarray}

Thus if we combine \eqref{61}, \eqref{59} and \eqref{54} along with $E^{\{2\}}_{\alpha,N-1}=E^{\{1\}}_{N+\alpha,0}$, we conclude that

\begin{eqnarray}\label{62}
\widetilde{H}^{(N-1)}_{\alpha+N} =H_{\alpha}
-2\frac{d^2}{dx^2} \ln \Ph_{\alpha,N-1} =  \widehat{H}^{(1)}_\alpha\,,
\end{eqnarray}

which is the desired result.

A final remark: The number $N$ should be odd, since only if $N-1$ is even, the function $\Ph_{\alpha,N-1}$ is nodeless.

\sect{Concluding remarks}

We have classified the poles of the scattering matrix of the Rosen--Morse II Hamiltonian. Depending on certain values of the parameters,  $\lambda$ and $ \beta$, from which the potential depends and which affect to its shape, this Hamiltonian may or may not have bound states. There are always two  infinite sequences  of anti-bound, or virtual, poles (referred to as type 1 and 2 poles) and, eventually a finite number of redundant states, which may also be of two types. This depends   on the values of the parameters. There are no resonance poles. 

We have studied the SUSY transformation of these Hamiltonians based on the states associated to the above mentioned poles.

If bound states existed, we may use the ground state wave function as a seed for a series of SUSY transformations, which generates the shape invariant hierarchy of Rosen--Morse II Hamiltonians
$H_{\lambda +n}$. We observe that each transformation erases  a bound state up to a finite number of them, when the resulting and successive Hamiltonians have no bound states. At the same time one antibound state of type 2 is gained with the same energy at each step. In this sense, the bound-antibound of type 2 behaviour under this kind of SUSY transformations is complementary. The number of redundant poles, whenever exists, remains the same. 
The connection of different Hamiltonians in the hierarchy $H_{\lambda +n}$ is realized by the shift operators $B^\pm_{\lambda  + n}$,  where $n$ is any integer. The problem of the building of ladder operators is not easy; it has been considered in \cite{hussin19,hussin21}.

If the seed function is a solution of the Schr\"odinger equation with energies given by redundant poles of $S(E)$, the resulting first--order transformation gives a Hamiltonian with the same number of poles of each kind. Both Hamiltonians are isospectral with respect to bound as well as to antibound states.

If the seed function is a solution with energy in a anti-bound state, the transformed Hamiltonian $\tilde H_\lambda$ has one more bound state while one antibound less than
$ H_\lambda$.

Finally, we have shown the equivalence, in terms of the final resulting Hamiltonian, between a SUSY transformation using an anti-bound state wave function and an even number of transformations using the ground state as a seed function, whenever bound states exist for the original Rosen--Morse II Hamiltonian.

In all the aforementioned SUSY transformations of Rosen--Morse II Hamiltonians based on the poles of the $S$ matrix give rise to rational partner potentials \cite{QUESNE}. Our results  show the interest of the states associated to such singularities; they generate SUSY transformations
which constitute a complement to the well known cases based only on bound states. We think that further research on applications of $S$ poles in SUSY transformations will
explain many other properties, for instance, related with exceptional polynomials \cite{IAN12}, the interaction of magnons-skyrmions \cite{LEE23}, the dynamics of perturbed nonrotating black holes \cite{LENCI23} etc. Other applications to graphene under magnetic fields have been considered in a  number of references \cite{zali,costescu,kuru09,david,david21,moran}.

\section*{Acknowledgements}

This work was  supported by MCIN, Spain with funding from the European Union
NexGenerationEU (PRTRC17.I1) and the Consejer\'ia de Educaci\'on de la Junta de Castilla y Le\'on, Spain, Project QCAYLE and the MCIN Project PID2020-113406GB-I00 of Spain.
\\
\\
Data Availability Statement: No Data associated in the manuscript.

\bigskip

\bibliographystyle{plain}

\end{document}